\documentstyle{article}
\begin{document}
\title{THEORY OF PUMP DEPLETION AND SPIKE FORMATION
       IN STIMULATED RAMAN SCATTERING}
\author{C. CLAUDE and J. LEON\\
Physique Math\'ematique et Th\'eorique, CNRS-URA 768\\
Universit\'e Montpellier II, 34095 MONTPELLIER FRANCE}
\date{}
\maketitle\begin{abstract}
By using the inverse spectral transform, the SRS equations are solved
and the explicit output data is given for arbitrary laser pump
and Stokes seed profiles injected on a vacuum of optical phonons.
For long duration laser pulses, this solution is modified such
as to take into account the damping rate of the optical phonon wave.
This model is used to interprete the experiments of
Dr\"uhl, Wenzel and  Carlsten (Phys. Rev. Lett., (1983) {\bf51}, 1171),
in particular the creation of a spike of  (anomalous) pump radiation.
The related nonlinear Fourier spectrum does not contain discrete
eigenvalue, hence this {\em Raman spike} is not a soliton.
\end{abstract}
\vskip6cm
\noindent\hrule
\vskip10pt
PACS \#  42.65Dr, 42.50 Rh,\hfill Preprint \#PM 94-16
\vskip5pt
\centerline{To appear in {\em Physical Review Letters}}
\pagebreak

Stimulated Raman scattering (SRS) is one of the
most studied three-wave interaction processes
in nonlinear optics not only because it retains all the
ingredients of any other stimulated process, but also because it has
revealed many striking, and sometimes unexplained phenomena.
The theory of SRS has been developped
on a semi-classical basis, for instance by  Shen and
Bloembergen \cite{shbloe}, Wang \cite{wang} or  Carman {\em et al}
\cite{carshi} by assuming a permanent pump intensity
profile. But when {\em "depletion of the
laser power implies that the laser field may not be treated as
a fixed constant parameter"} \cite{shbloe}, Stokes generation and
amplification  induces pump  depletion and it is a
serious obstacle to the propagation of high intensity
laser pulses in a Raman-active medium.
However it can also be thought of as a means to study
(experimentaly and theoreticaly) the fundamental properties
of matter and radiation, and
indeed Raymer and Mostowski \cite{raymost}  predict large (macroscopic)
fluctuations of the Stokes pulse energy that are reminiscent of the
small (quantum) fluctuations of the material dynamical variable
(the polarization density state variable).
These predictions were then cheked on measurements of the statistical
distribution of Stokes pulses in pressured $H_2$ gas by Wamsley and
Raymer \cite{wamray} and also by Fabricius, Natterman and von der Linde
\cite{fabnat}.

At the same time, experiments of Dr\"uhl, Wenzel and  Carlsten
\cite{druwen} for long duration pump pulses (order of 100ns)
revealed that {\em "ocasionally the pump depletion is
anomalously reversed for a short time interval generating a spike of
pump radiation"}. They call it a
{\em soliton} refering to the soliton-solution of the
undamped SRS equations given by Chu and Scott \cite{chusco}. The
lesson of these results is the spectacular fit of the
experimental data with numerical simulations of the SRS equations,
indicating that the model is quite adequate. Then the fundamental result
is the discovery that the {\em spike} in the pump-depletion
occurs when a $\pi$-phase shift is introduced in the Stokes seed.

In the absence of phase flip, the spike formation is still obseved
(on real experiments, not on simulations)  but with much lower probability.
This led  Englund and Bowden \cite{engbow} to propose that the spike  is the
macroscopic manisfestation of quantum {\em phase fluctuations} of the Stokes
wave. The subsequent experiments of MacPherson, Swanson and Carlsten
revealed that the anomalous  pump
radiation spike (what they call a {\em Raman soliton}) occurs in about
10.1\% of the shots \cite{macswa}. Englund and Bowden \cite{engbow2}
developed a complete theoretical basis of such Raman spikes generated by
quantum fluctuations of the initial Stokes vacuum, and they obtained a
reasonable qualitative agreement with the previous experiments (they
found a spontaneously generated spike in 13.6\% of the shots).

The observed spikes of pump radiation acquire then importance also for
fundamental studies and they are always refered to as being {\em solitons}
\cite{soliton}. But it was already remarked on the first series of
numerical simulations  that the spike narrowing
(as propagation distance is increased) indicates that
they are {\em "not solitary waves in the strict sense"} \cite{druwen}.
In particular the non-zero velocity of the {\em Raman soliton} implies,
as shown by Menyuk, that {\em "any solitonlike structures are subluminous
and will ultimately  disappear at the back end of the pulse"} \cite{menyuk}.

Consequently the problem of the theoretical interpretation of the experiments
of Dr\"uhl, Wenzel and  Carlsten \cite{druwen} is still open and we solve it
here in terms of the inverse spectral transform theory (IST) extended to
{\em arbitrary boundary values} \cite{leon}. We obatain the {\em explicit
global} solution  (output laser pulse) which maps perfectly the numerical
simulation of SRS (as performed by MacPherson, Carlsten and Druhl
\cite{macarl}).

Such an  {\em explicit analytic formula} for the output (eqs. (\ref{outa1})
and (\ref{outa2}) below) is important for physics in many aspects:\\
1 - It allows to understand why the Stokes phase flip and the finiteness
of the dephasing time (long pulses) are {\em both} essential to the spike
formation, and to discover the precise nonlinear mechanism generating
the Raman spike.\\
2 - It provides a powerfull tool to analyse the experiments. Indeed,
having the digitalized input, our formula readily gives the predicted output
and a comparison with the data of \cite{druwen} will be published later with
other details \cite{pra-srs}.\\
3 - It unveils the nature of the Raman spike as its {\em
nonlinear Fourier spectrum} does not contain isolated points (bound states)
but consists only in the contimuum (radiation). Apart form the fact
that the spike is not a soliton, the important consequence of this is that it
survives long propagation distances, which is important for applications.

The last point to mention is the fact that our solution, although being
obatined from the {\em infinite line case}, is quite close to the finite line
case solution (obtained through numerical simulations). This is easily be
seen by comparing for instance our figure 2 to the figure 2 of \cite{macarl}.
The reason for this is that the input data are of {\em finite duration}
and the interaction is {\em local} (for any fixed time). We show in
\cite{pra-srs} that indeed the generated material excitation is localized in
a small region (of length comparable with the duration of the amplified
Stokes pulse).

The model of SRS can be taken for instance from \cite{engbow2}
and reads, if we neglect the ground state depletion ($R_3= -N$):
$$
\partial_\zeta A_L=K_{LS} R A_S,\hskip10pt
\partial_\zeta A_S=-K_{LS}^* R^* A_L $$
\begin{equation}
\partial_\tau R+RL/(cT_2) =-K_{LS} N A_L A_S^*.
\label{srsphys}\end{equation}
The spatial variable $\zeta$ lies in $[0,L]$, where $L$ is the total
beam path in the Raman cell, and the retarded time $\tau=t-\zeta/c$
is positive.
$A_L$ and $A_S$ are the slow envelopes of the laser pump (frequency
$\omega_L$) and of the Stokes emission (frequency $\omega_S$) which
stimulate the material dynamical variable $R$ (optical
phonon, frequency $\omega_P=\omega_L-\omega_S$).
$K_{LS}$ is the complex coupling
constant, $T_2$ the mean collisional dephasing time and $N$ is a scaled
density ($=\rho AL$ with $\rho$ the density of Raman active molecules
and $A$ the effective crossectional area of the pump beam).
The initial-boundary value problem associated
to the system (\ref{srsphys}) is the following:
\begin{equation}
R(\zeta,0)=0,\hskip10pt A_L(0,\tau)=A_{L0}(\tau),\hskip10pt
 A_S(0,\tau)=A_{S0}(\tau),
\label{inbophys}\end{equation}
where, to reproduce the experiments, $A_{L0}$ is a gaussian,
$A_{S0}$ is a fraction of $A_{L0}$ with possibly a change of sign (phase
flip) somewhere. The problem is to determine
the output quantities $A_L(L,\tau)$ and $A_S(L,\tau)$.

Our model results from (\ref{srsphys}) by considering first an infinite
line ($L\to\infty$) and taking into account the mismatch wave number
of value $2k=k_P+k_S-k_L$, which results for instance from the D\"oppler
effect due to molecular thermal motions. The resulting system is
(\ref{srsphys}) but with $A_S$ replaced with
$A_S\exp[-2ik\zeta]$. To take into account the contributions of all
values of $k$, we introduce the distribution $g(k)$ (centered in $k=0$)
of the relative coupling intensities. The resulting model equations are
$$
\partial_x a_1=q a_2,\hskip10pt
\partial_x a_2-2ika_2=-\bar q a_1 $$
\begin{equation}
\partial_t q+\gamma q=\int dk g(k)a_1\bar a_2.
\label{srsmath}\end{equation}
Here and in the following, an integral with no specified boundaries stands
for $(-\infty,+\infty)$.
Hereabove we have made the following change of variables and scalings:
\begin{equation}
x=-\zeta,\hskip10pt t=\tau,\hskip10pt q=-K_{LS}R,\hskip10pt
\gamma=L/(cT_2),
\label{spacevar}\end{equation}
\begin{equation}
(k=0):\hskip10pt a_1=A_L/A_0,\hskip10pt
a_2=A_Se^{2ikx}/A_0,
\label{fieldvar}\end{equation}
where $A_0=Max\{ A_L(0,\tau)\}$. The distribution $g(k)$ is actually related
to the {\em inhomogeneous broadening} and can be
normalized to the coupling constant by setting
$\int dk g(k)=K_{LS}^2N|A_0|^2$.
The related initial-boundary value problem corresponding to
(\ref{inbophys}) reads here (note the sign in (\ref{spacevar}))
\begin{equation}
q(x,0)=0,\hskip10pt a_1(k,+\infty,t)=I_1(k,t),\hskip10pt
a_2(k,+\infty,t)=I_2(k,t)e^{2ikx}
\label{inbomath}\end{equation}
where for $k=0$, $I_1(t)=A_{L0}(\tau)/A_0$ and $I_2(t)=A_{S0}(\tau)/A_0$.
The problem to be solved is now to compute the output
data $a_1(k,-\infty,t)$ and $a_2(k,-\infty,t)$.

Note that, although $t$ represents the retarded time, the initial value problem
is physically meaningful because, the medium being initially in the ground
state, we have set $q(x,0)=0$. Another important remark is that the model
(\ref{srsmath}) maps onto (\ref{srsphys}) in the limit when $g(k)$ becomes
the Dirac distribution $\delta(k)$. However, on the infinite line, this is
a singular limit \cite{leon} (in short it is not compatible with
$q(x,t)\to 0$ as $x\to\pm\infty$), and hence $g(k)$ can be as sharp as we
want but never a true delta function (which is physically quite
reasonable).

It has been shown in \cite{leon} that the above initial-boundary value
problem is solvable for $\gamma=0$ and we shall use here directly these
results. The solution is given by
\begin{equation}
a_1(k,-\infty,t)=I_1/\bar\beta+I_2\bar\alpha/\bar\beta,
\label{outa1}\end{equation}
\begin{equation}
a_2(k,-\infty,t)e^{-2ikx}=I_2/\beta-I_1\alpha/\beta.
\label{outa2}\end{equation}
The coefficients $\alpha(k,t)$ and $\beta(k,t)$ (the {\em spectral data})
can be computed explicitely from \cite{leon}  and read for $\gamma=0$
\begin{equation}
\alpha(k,t)=-\pi g(k)e^{\phi(k,t)}\int_{0}^{t}dt'\bar I_1(k,t')I_2(k,t')
e^{-\phi(k,t')}
\label{alpha}\end{equation}
\begin{equation}
\phi(k,t)=\int_{0}^{t}dt'\left[{1\over2}\pi g(k) U(k,t')-{i\over2}
\int{\!\!\!\!\!\! -}
{d\lambda\over\lambda-k} g(\lambda) U(\lambda,t')\right]
\label{phi}\end{equation}
\begin{equation}
U(k,t)=|I_1(k,t)|^2-|I_2(k,t)|^2
\label{u}\end{equation}
\begin{equation}
\beta(k,t)=\sqrt{1+|\alpha(k,t)|^2}\hskip5pt e^{i\theta(k,t)}
\label{beta}\end{equation}
\begin{equation}
\theta(k,t)=-{1\over2\pi}
\int{\!\!\!\!\!\! -}
{d\lambda\over\lambda-k}\log(1+|\alpha(\lambda,t)|^2),
\label{theta}\end{equation}
where the slashed integral denotes the Cauchy principal value.
Analogous formulae can be found in \cite{gabzak} but in the context of
resonant interaction of light with a two-level medium (and application to
superfluorescence) for which the boundary value problem notably differs.
The above result
gives the exact solution to SRS for short pump pulses (for which
$\gamma\sim0$). We report the discussion of this case to forthcoming
paper and consider now the case of long pump pulses for which the pump
depletion can be anomalously reversed \cite{druwen}.

Our main argument is that both pump depletion and spike formation are
described by the above solution of the boundary-value problem
(\ref{inbomath}) for the system (\ref{srsmath}). Indeed,
considering (\ref{outa1}) and (\ref{beta}), we remark that
\begin{equation}
\alpha \to \infty\hskip5pt\Rightarrow\hskip5pt|\beta|\to\infty
\hskip5pt\Rightarrow\hskip5pt|a_1(-\infty)|\to|I_2|,
\label{pumpdep}\end{equation}
\begin{equation}
\alpha \to 0\hskip5pt\Rightarrow\hskip5pt|\beta|\to 1
\hskip5pt\Rightarrow\hskip5pt|a_1(-\infty)|\to|I_1|.
\label{spike}\end{equation}
Then pump depletion will occur in the time region where $\alpha(k,t)$ is
large (the pump output will be of the order of the Stokes {\em input}
$I_2$), and pump radiation will occur
in the time region where $\alpha(k,t)$ is close to zero
(the pump output will be of the order of the pump {\em input} $I_1$).

In the case of long duration pump pulses,
the effect of the dephasing time is included in our model by assuming
instead of the evolution (\ref{alpha}) the following one
\begin{equation}
\alpha(k,t)=-\pi g(k)e^{\phi(k,t)}\int_{0}^{t}dt' \bar I_1(k,t')I_2(k,t')
e^{-\phi(k,t')}e^{\gamma(t'-t)}.
\label{alphadamp}\end{equation}
The added exponential factor hereabove is justified by considering
the linear limit in the spectral transform context. To that end it is
convenient to write down the solution of (\ref{srsmath}) given by the
spectral transform method \cite{leon}:
\begin{equation}
a_1=I_1f_1+I_2e^{2ikx}f_2,
\label{a1linear}\end{equation}
\begin{equation}
a_2=-I_1\bar f_2+I_2e^{2ikx}\bar f_1,
\label{a2linear}\end{equation}
\begin{equation}
q=-{1\over\pi}\int dk\;\;\bar\alpha e^{-2ikx} f_1,
\label{qlinear}\end{equation}
where $f_1$ and  $f_2$ are the solution of
$$
f_1(k,x,t)=1+{1\over2i\pi}\int{d\lambda\over\lambda+i0-k}
\alpha(\lambda,t)e^{2i\lambda x}f_2(\lambda,x,t),
$$
\begin{equation}
f_2(k,x,t)={1\over2i\pi}\int{d\lambda\over\lambda-i0-k}
\bar\alpha(\lambda,t)e^{-2i\lambda x}f_1(\lambda,x,t).
\end{equation}
With the linear limit, obtained from the above solution by taking
simply $f_1=1$, it can be verified that the evolution for $q$
in (\ref{srsmath}) with $\gamma\ne0$ has precisely the solution
(\ref{alphadamp}). We remark at this stage that the time evolution
of the nonlinear Fourier transform $\alpha(k,t)$ bears the {\em linear}
character of the evolution. Hence the nonlinearity enters only in the
output expressions (\ref{outa1}) and (\ref{outa2}) through the nonlinear
combinations of $\alpha$ with $I_{1,2}$.

In order to realize how the mechanism of pump depletion and
spike formation is allowed by the equation (\ref{alphadamp}),
we deal with all quantities evaluated
at $k=0$ (corresponding to a very sharp $g(k)$). Then we assume
a Gaussian shaped laser pump input
(from (\ref{fieldvar}) the amplitude is normalized to 1)
and a small proportion of Stokes seed with possibly a $\pi$-phase shift:
\begin{equation}
I_1(t)=\exp[-(t-t_1)^2/\tau_1^2],\hskip10pt
I_2(t)=\tanh[(t-t_0)/\tau_0]I_1(t)/\Gamma.
\label{inputs}\end{equation}
With such input data, the pump repletion can be physically understood as
resulting
from a reversal of the Raman gain due to the change of sign of the Stokes
input.
This is precisely this behavior which is described by the formula
(\ref{alphadamp})
where, if $I_2$ changes sign, then $\alpha$ starts to decrease and the
pump repletion limit (\ref{spike}) is approaches.

We have drawn the energy $|a_1|^2$ of
the pump output given by (\ref{outa1}) on figs. 1 and 2, for the
above choices of $I_{1,2}$ with the  parameter values
$t_1=50,\hskip5pt \tau_1=45.5,\hskip5pt \Gamma=10,\hskip5pt
g(0)=100,\hskip5pt \gamma=160 $
and no Stokes phase flip in fig 1 (that is $t_0 =0$) and
a phase flip in $t_0=50$ for the fig 2 (with $\tau_0=1$). These parameters
correspond in the physical world to a pump pulse of maximum amplitude $A_0$ of
$6.36\ 10^6\ Vm^{-1}$ and a Stokes seed of $0.636\ 10^6\ Vm^{-1}$,
for the values (taken from ref. \cite{engbow2})
for $\tau$ in nanosec: $N=8.5\ 10^{19}$,  $|K_{LS}|=5.8\ 10^{-17}$ and
$T_2=0.625\ ns$. Consequently, from $\gamma=L/(cT_2)=160$ the data would
correspond to a beam path of $30\ m$.

\hspace{-2cm}
%
\setlength{\unitlength}{0.240900pt}
\ifx\plotpoint\undefined\newsavebox{\plotpoint}\fi
\sbox{\plotpoint}{\rule[-0.175pt]{0.350pt}{0.350pt}}%
\\
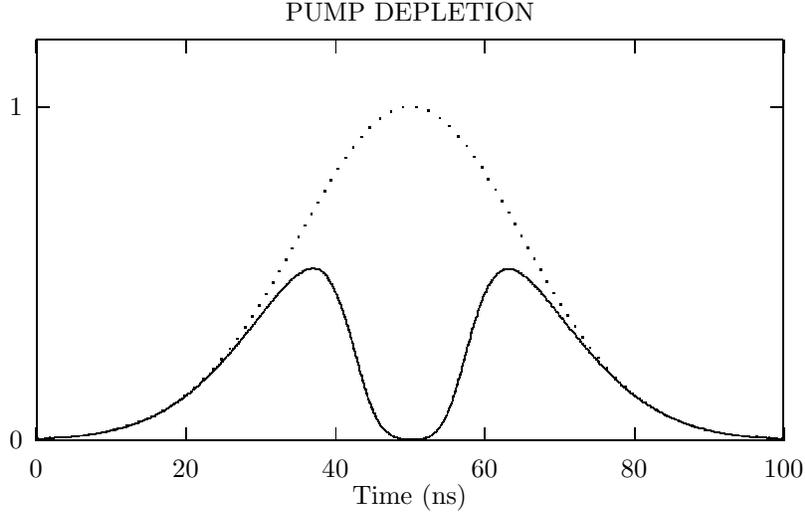
{\em Fig.1: Pump energy profile $|a_1|^2$ at $x=+\infty$ (input, dashed
line) and at $x=-\infty$ (output, solid line) with unflipped phase of
the Stokes seed.}

\hspace{-2cm}
%
%
\setlength{\unitlength}{0.240900pt}
\ifx\plotpoint\undefined\newsavebox{\plotpoint}\fi
\sbox{\plotpoint}{\rule[-0.175pt]{0.350pt}{0.350pt}}%
%
\\
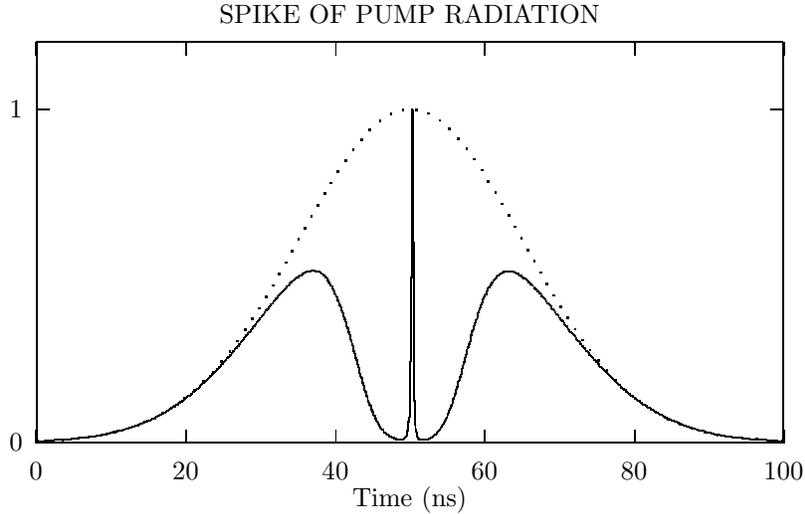
{\em Fig.2: Pump energy profile $|a_1|^2$ at $x=+\infty$ (input, dashed
line) and at $x=-\infty$ (output, solid line) when a $\pi$-phase shift of
the Stokes seed is introduced at $50ns$.}

We can now easily understand the different behaviors of the pump output on
figs. 1 and 2 just by inspection of (\ref{alphadamp}). Indeed, in the
first zone (up to $30ns$), $\alpha$ is small and we have no depletion.
Then the growth of $\alpha$ (pump depletion) results from the
factor $\exp[\phi]$ in (\ref{alphadamp}) because $U$ given by
(\ref{u}) is {\em positive}. Now if $I_2$ in (\ref{u})
does not change sign (fig 1),  $\alpha$ grows up to when the damping term
dominates again and we observe the pump radiation again (right hand
side hump in fig 1 and 2).
If instead we introduce a phase flip in the Stokes seed, then the
integral in  (\ref{u}) makes $\alpha$ to decrease and possibly to
vanish, which constitutes the mechanism for reversal of pump depletion.
However this is allowed only if the growth of $\alpha$ with $\exp[\phi]$
is not too fast, and here enters the role of the damping term
$\exp[-\gamma t]$. Hence the spike of pump radiation occurs as a result
of Stokes phase flip via a balance between Stokes amplification
and optical phonon damping.

In conclusion, we have obtained the following set of results:\\
1 - An exact and explicit solution to the SRS equations for short pulses
($\gamma=0$).\\
2 - An approximate solution for long pump pulses which describes perfectly
in a unique formalism the pump depletion and the formation
of a spike of pump radiation when the Stokes seed is given a phase flip.\\
3 - The proof that the spike of pump radiation occurs as a balance between
Raman gain and phonon damping, and that it can survive long
propagation distances.\\
4 - A new mathematical structure, the {\em Raman spike}, related to a zero of
the reflection coefficient (our $\alpha(k,t)$), and which, in the IST scheme,
is part of the continuous spectrum.\\
5 - The proof that the spike of pump radiation is not a soliton, namely
that it is not related to a discrete part of the (nonlinear Fourier)
spectrum.\\
6 - An explicit formula for the description of transient SRS which can be used
to study for instance the decay of the Raman spike, the generation of
multi-spikes, the result of a stochastic phase in the generated Stokes
wave, etc... These studies will be reported elsewhere (see \cite{pra-srs}).

\end{document}